\documentclass[12pt]{article}
\usepackage{amsmath,amsthm,epsfig,graphicx,amssymb,amsfonts}

\newcommand{\Qx}{ \mathbb{Q} }

\newcommand{\cds}{\mbox{CDS}}

\newcommand{\pcds}{\Pi\mbox{\tiny RCDS}}

\newcommand{\npv}{\mbox{NPV}}

\newcommand{\lgd}{\mbox{L{\tiny GD}}}
\newcommand{\rec}{\mbox{R{\tiny EC}}}

\newtheorem{theorem}{Theorem}[section]
\newtheorem{proposition}[theorem]{Proposition}

\title{\vspace{-2cm} {\small  An extended and updated version of this paper with the title} \\
 {\small \bf Credit Calibration with Structural Models} \\{\small \bf and Equity Return Swap valuation under Counterparty Risk} \\ {\small \bf will appear in:} {\small Bielecki, Brigo and Patras (Editors)}\\ {\small  Recent advancements in theory and practice of credit derivatives, Bloomberg Press, 2010.}\\ --- \\ \bf{\large  Credit Calibration with Structural Models: The Lehman case and Equity Swaps under Counterparty Risk}}
\author{
Damiano Brigo\thanks{Dept. of Mathematics, Imperial College, and Fitch Solutions. {\tt d.brigo@imperial.ac.uk}}  \ \ \ Massimo Morini\thanks{Banca IMI and Bocconi University, Milan, {\tt massimo.morini@bancaimi.com }} \ \ \ \ Marco Tarenghi\thanks{Banca Leonardo, Milan, {\tt marco.tarenghi@bancaleonardo.com}} \\
}
\date{\small First version: November 9, 2009. This Version: \today}

\begin{document}

\maketitle \thispagestyle{empty}

\vspace{-0.75cm}
\begin{abstract}
In this paper we develop structural first passage models (AT1P and SBTV) with time-varying volatility and characterized by high tractability, moving from the original work of Brigo and Tarenghi (2004,  2005) \cite{BrTar04} \cite{BrTar05} and Brigo and Morini (2006)\cite{BrMor}. The models can be calibrated exactly to credit spreads using efficient closed-form formulas for default probabilities. Default events are caused by the value of the firm assets hitting a safety threshold, which depends on the financial situation of the company and on market conditions. In AT1P this default barrier is deterministic. Instead SBTV assumes two possible scenarios for the initial level of the default barrier, for taking into account uncertainty on balance sheet information. While in \cite{BrTar04} and \cite{BrMor} the models are analyzed across Parmalat's history, here we apply the models to exact calibration of Lehman Credit Default Swap (CDS) data during the months preceding default, as the crisis unfolds. The results we obtain with AT1P and SBTV have reasonable economic interpretation, and are particularly realistic when SBTV is considered. The pricing of counterparty risk in an Equity Return Swap is a convenient application we consider, also to illustrate the interaction of our credit models with equity models in hybrid products context.
\end{abstract}

\subsection*{Keywords}
{\small Credit Default Swaps, Structural Models, Black Cox Model, Calibration,
Analytical Tractability, Monte Carlo Simulation, Equity Return Swaps, Counterparty
Risk, Barrier Options, Uncertain Credit Quality, Lehman Brothers Default.}

\section{Introduction}

Modeling firms default is an important issue, especially in
recent times where the crisis begun in 2007 has led to bankruptcy of several companies all over the world. 
\emph{Structural models} are based on the work by Merton (1974),
in which a firm life is linked to its ability to pay back its
debt. Let us suppose that a firm issues a bond to finance its
activities and also that this bond has maturity $T$. At final time
$T$, if the firm is not able to reimburse all the bondholders we
can say that there has been a default event. In this simplified context
default may occur only at final time $T$ and is triggered by the
value of the firm being below the debt level. In a more realistic
and sophisticated structural model (Black and Cox (BC) (1976),
part of the family of \emph{first passage time models}) default
can happen also before maturity $T$. In first passage time models
the default time is the first instant where the firm value hits
from above either a deterministic (possibly time varying) or a
stochastic barrier, ideally associated with safety covenants
forcing the firm to early bankruptcy in case of important credit
deterioration. In this sense the firm value is seen as a generic
asset and these models use the same mathematics of barrier options
pricing models. For a summary of the literature on structural
models, possibly with stochastic interest rates and default
barriers, we refer for example to Chapter 3 of Bielecki and
Rutkowski (2001). Here we just mention works
generalizing the form of the default barrier to account for the debt
level, such as Briys and de Varenne (1997), Hsu et al. (2002), Hui et al. (2003), and
Dorfleitner, Schneider and Veza (2007).

It is important to notice that structural models
make some implicit but important assumptions: They assume that the
firm value follows a random process similar to the one used to
describe generic stocks in equity markets, and that it is possible
to observe this value or to obtain it from equity information. Therefore, unlike intensity
models, here the default process can be completely monitored based
on default free market information and comes less as a surprise.
However, structural models in their basic formulations and with
standard barriers (Merton, BC) have few parameters in their
dynamics and cannot be calibrated exactly to structured data such
as CDS quotes along different maturities.

In this paper we follow \cite{BrTar04}, \cite{BrTar05}, \cite{BrMor} and \cite{BrMorTar}
 to effectively use two families of structural
models (AT1P and SBTV) in the ideal ``territory" of intensity models,
i.e. to describe default probabilities in a way that is rich enough
to calibrate CDS quotes. First of all, in Section~\ref{sec:at1p}
we present the Analytically Tractable First Passage Model (AT1P),
which has a deterministic default barrier and is a generalization
of the classic Black and Cox model; we show how this model, differently
from the Black and Cox Model, mantains analytical formulas for
default probabilities even when allowing asset volatility to be
realistically time-varying. In Section~\ref{sec:calibration} we
show how this feature can be exploited in order to calibrate the
parameters of the model to market prices of single name CDS's (in
a way similar to the procedure used to calibrate intensities in
reduced form models). While in \cite{BrTar04} and \cite{BrMor} we
had tested AT1P and SBTV calibration on Parmalat data as Parmalat approached default,
here in Section~\ref{sec:lehmanat1p} we consider
the concrete case of another firm, Lehman Brothers, approaching default,
and we show how our proposed structural model calibration changes
as the company credit quality (as summarized by its CDS quotes)
deteriorates in time. The analysis of the model behaviour in
representing Lehman approaching default leads us to introduce
in Section~\ref{sec:sbtv} an extension of the model, aimed at
adding further realism to the model. This extension,
called Scenario Barrier Time-Varying Volatility Model (SBTV),
assumes two possible scenarios for the initial level of the
default barrier, to take into account uncertainty on balance sheet
information, and in particular the difficulty in evaluating the
liability portfolio of the company. Also in this case we mantain
analytical tractability and in Section~\ref{sec:lehmansbtv} we present
the results of calibration to CDS quotes.

Due to their intrinsic nature, structural models result to be a natural
method to price equity-credit hybrid products and also to evaluate
counterparty risk in equity products. The counterparty
risk (Credit Valuation) Adjustment (CVA) has received a lot of attention in
an exploding literature, and in several asset classes. We recall here the work of
Sorensen and Bollier (1994), Huge and Lando (1999), Brigo and Pallavicini (2007),
Brigo and Masetti (2005) and Brigo, Pallavicini and Papatheodorou (2009) for CVA
on interest rate products, both unilateral and bilateral (last paper). For credit
products, especially CDS, CVA has been analyzed in Leung and Kwok (2005),
Blanchet-Scalliet and Patras (2008), Brigo and Chourdakis (2009), Crepey,
Jeanblanc and Zargari (2009), Lipton and Sepp (2009), who resort to a structural
model, and Walker (2005), while Brigo and Capponi (2008) study the bilateral case.
CVA on commodities is analyzed in Brigo and Bakkar (2009), whereas CVA on Equity is
analyzed in Brigo and Tarenghi (2004, 2005). CVA with netting is examined in
Brigo and Masetti (2005), Brigo and Pallavicini (2007) and Brigo, Pallavicini and
Papatheodorou (2009). CVA with collateral margining is analyzed in Assefa, Bielecki,
Crepey and Jeanblanc (2009) and in Alavian et al (2009), whereas collateral triggers are considered in Yi (2009).
In Section~\ref{sec:otherproducts} we consider the particular case of an Equity Return Swap with counterparty
risk. In this context we have to take care of the correlation between
the counterparty and the underlying, and this is done much more
conveniently in structural models than in intensity models. Here for CVA results in equity payoffs we follow
Brigo and Tarenghi (2004, 2005), also reported in the final part of Brigo and Masetti (2005). 
To help the reader orientation with respect our past works on CVA, we summarize them in Table~\ref{table:contextCRmodels}. This is a table that is supposed to help putting our paper in the context of our earlier works on CVA and is not meant to be exhaustive or even representative of the research field.

\begin{table}
\begin{center}
{\small
\begin{tabular}{|c||c|c|c|}
\hline
 CVA & \multicolumn{ 2}{|c|}{One-sided} &  Bilateral \\\hline
Modelling & Volatility & \multicolumn{ 2}{|c|}{Volatility and Correlation} \\\hline\hline
IR Swaps & Brigo Masetti (2005)  & Brigo Pallavicini (2007) & Brigo, Pallavicini and  \\
{\scriptsize with Netting} &  & &  Papatheodorou (2009) \\\hline
IR Exotics &            & Brigo Pallavicini (2007) & B., P. and P. (2009)   \\\hline
 Oil Swaps &            & Brigo Bakkar (2009) &            \\\hline
       CDS &            & Brigo Chourdakis (2008) & Brigo Capponi (2008) \\ \hline
Equity TRS &            & Brigo Tarenghi (2004,2005) &            \\
 &            & Brigo Masetti (2005) &            \\\hline
\end{tabular}}
\end{center}
\caption{Part of earlier analogous literature on CVA valuation with respect to inclusion of underlying asset volatilities and/or correlation between underlying asset and counterparties, along with bilateral features.}\label{table:contextCRmodels}
\end{table}

\section{The Analytically Tractable First Passage (AT1P) Model}\label{sec:at1p}

The fundamental hypothesis of the model we resume here is that the
underlying process is a Geometric Brownian Motion (GBM), which is
also the kind of process commonly used for equity stocks in the
Black Scholes model. This choice is the typical choice of classical
structural models (Merton, Black Cox), which postulate a GBM
(Black and Scholes) lognormal dynamics for the value of the firm
$V$. This lognormality assumption is considered to be acceptable.
Crouhy et al (2000) report that ``this assumption [lognormal V] is
quite robust and, according to KMV's own empirical studies, actual
data conform quite well to this hypothesis.". 

\begin{proposition}\label{prop:at1p} {\bf (Analytically-Tractable
First Passage (AT1P) Model)} Assume the risk neutral dynamics for
the value of the firm $V$ is characterized by a risk free rate
$r_t$, a payout ratio $k_t$ and an instantaneous volatility
$\sigma_t$, according to equation
\begin{equation}\label{valueproc} dV_t = V_t\,(r_t-k_t)\,dt+V_t\,\sigma_t\,dW_t
\end{equation}
and assume a
default barrier $H(t)$ (depending on the parameters $H$ and $B$) of the form
\[ H(t) = H\exp\left(\int_0^t\left(r_u-k_u-B\sigma^2_u\right)du\right)\] and let $\tau$ be defined
as the first time where $V(t)$ hits $H(t)$ from above, starting from
$V_0>H$,
\[ \tau = \inf\{ t \ge 0: V_t \le H(t)\}. \]
Then the survival probability is given analytically by
\begin{eqnarray}\label{survcoher}
\mathbb{Q}\left(\tau> T \right) = \left[\Phi\left(\frac{\log \frac{V_0}{H}
+\frac{2B-1}{2}\int_0^T \sigma^2_u du}{\sqrt{\int_0^T\sigma^2_u du}}\right)-
\left(\frac{H}{V_0}\right)^{2B-1}\Phi\left(\frac{\log
\frac{H}{V_0} +\frac{2B-1}{2}\int_0^T \sigma^2_u
du}{\sqrt{\int_0^T\sigma^2_u du}}\right)\right].
\end{eqnarray}
\end{proposition}
For a proof, see \cite{BrTar04} or \cite{BrMorTar}. 

A couple of remarks are in order. First of all, we notice that in the formula for the survival probability in Proposition \ref{prop:at1p}, $H$ and $V$ never appear alone, but always in ratios like $V/H$; this homogeneity property allows us to rescale the initial value of the firm $V_0=1$, and express the barrier parameter $H$ as a fraction of it. In this case we do not need to know the real value of the firm, neither its real debt situation. Also, we can re-write the barrier as
\begin{eqnarray}\label{barrier}
\nonumber H(t) 
 =  \frac{H}{V_0}\mathbb{E}_0\left[V_t\right]\exp\left(-B\int_0^t\sigma^2_udu\right).
\end{eqnarray}
Therefore the behaviour of $H(t)$ has a simple economic interpretation. The backbone of the default barrier at $t$ is a proportion, controlled by the parameter $H$, of the expected value of the company assets at $t$. $H$ may depend on the level of liabilities, on safety covenants and in general on the characteristics of the capital structure of the company. This is in line with observations in Giesecke (2004), pointing out that some discrepancies between the Black and Cox Model and empirical regularities may be addressed with the realistic assumption that, like the firm value, the total debt grows at a positive rate, or that firms maintain some target leverage ratio as in Collin-Dufresne and Goldstein (2001).

Also, depending on the value of the parameter $B$, it is possible that this backbone is modified by accounting for the volatility of the company's assets. For example, $B>0$ corresponds to the interpretation that when volatility increases -which can be independent of credit quality- the barrier is slightly lowered to cut some more slack to the company before going bankrupt. In the following tests we simply assume $B=0$, corresponding to a case where the barrier does not depend on volatility and the ``distance to default'' is simply modeled through the barrier parameter $H$.

\section{Calibration of the structural model to CDS data}\label{sec:calibration}
Since we are dealing with default probabilities of firms, it is
straightforward to think of financial instruments depending on
these probabilities and whose final aim is to protect against the
default event. One of the most representative protection
instruments is the Credit Default Swap (CDS). CDS's are contracts
that have been designed to offer protection against default. Here we introduce CDS in their traditional ``running'' form. For a methodology for converting running CDS to upfront CDS, as from the so called ``BIG BANG'' by ISDA, see Beumee et al. (2009).

Consider two companies ``A" (the \emph{protection buyer}) and ``B"
(the \emph{protection seller}) who agree on the following. If a third reference company ``C" (the \emph{reference credit})
defaults at a time $\tau_C \in (T_a,T_b]$, ``B" pays to ``A" at
time $\tau=\tau_C$ itself a certain ``protection" cash amount
$\lgd$ (Loss Given the Default of ``C" ), supposed to be
deterministic in the present paper. This cash amount is a {\em
protection} for ``A" in case ``C" defaults. A typical stylized
case occurs when ``A" has bought a corporate bond issued from ``C"
and is waiting for the coupons and final notional payment from
this bond: If ``C" defaults before the corporate bond maturity,
``A" does not receive such payments. ``A" then goes to ``B" and
buys some protection against this risk, asking ``B" a payment that
roughly amounts to the bond notional in case ``C" defaults.

Typically $\lgd$ is equal to a notional amount, or to a notional
amount minus a recovery rate. We denote the recovery rate by
``$\rec$".

In exchange for this protection, company ``A" agrees to pay
periodically to ``B" a fixed ``running" amount $R$, called ``CDS spread'', at a set of
times $\{T_{a+1},\ldots,T_b\}$,  $\alpha_i = T_{i}-T_{i-1}$,
$T_0=0$. These payments constitute the ``premium leg" of the CDS
(as opposed to the $\lgd$ payment, which is termed the
``protection leg"), and $R$ is fixed in advance at time $0$; the
premium payments go on up to default time $\tau$ if this occurs
before maturity $T_b$, or until maturity $T_b$ if no default
occurs.

\[ \begin{array}{ccccc} \mbox{``B"} & \rightarrow & \mbox{ protection } \lgd \mbox{ at default $\tau_C$ if $T_a< \tau_C \le T_b$} & \rightarrow & \mbox{``A"} \\
%
\mbox{``B"} & \leftarrow  & \mbox{ rate } R \mbox{ at }
T_{a+1},\ldots,T_b \mbox{ or until default } \tau_C & \leftarrow &
\mbox{``A"}
\end{array} \]

Formally, we may write the RCDS (``R" stands for running)
discounted value at time $t$ seen from ``A" as
\begin{eqnarray}\label{discountedpayoffcds}
\nonumber \pcds_{a,b}(t) := - D(t,\tau) (\tau-T_{\beta(\tau)-1}) R
\mathbf{1}_{\{T_a < \tau < T_b \} }
 -   \sum_{i=a+1}^b D(t,T_i) \alpha_i R \mathbf{1}_{\{\tau \ge T_i\}
 } \\
 + \mathbf{1}_{\{T_a < \tau \le T_b \} }D(t,\tau) \ \lgd
\end{eqnarray}
where $t\in [T_{\beta(t)-1},T_{\beta(t)})$, i.e. $T_{\beta(t)}$ is
the first date among the $T_i$'s that follows $t$, and where
$\alpha_i$ is the year fraction between $T_{i-1}$ and $T_i$.

The pricing formula for this
payoff depends on the assumptions on the interest rates dynamics
and on the default time $\tau$. Let $\mathcal{F}_t$ denote the
basic filtration without default, typically representing the
information flow of interest rates and possibly other default-free
market quantities (and also intensities in the case of reduced
form models), and $\mathcal{G}_t = \mathcal{F}_t\vee
\sigma\left(\{\tau<u\},u\leq t\right)$ the extended filtration
including explicit default information. In our current
``structural model" framework with deterministic default barrier
the two sigma-algebras coincide by construction, i.e.
$\mathcal{G}_t = \mathcal{F}_t$, because here the default is
completely driven by default-free market information. This is not
the case with intensity models, where the default is governed by
an external random variable and $\mathcal{F}_t$ is strictly
included in $\mathcal{G}_t$, i.e. $\mathcal{F}_t \subset
\mathcal{G}_t$.

We denote by $\cds(t,[T_{a+1},\ldots,T_b],T_a,T_b, R, \lgd)$ the
price at time $t$ of the above standard running CDS. At times some
terms are omitted, such as for example the list of payment dates
$[T_{a+1},\ldots,T_b]$. In general we can compute the CDS price
according to risk-neutral valuation (see for example Bielecki and
Rutkowski (2001)):
\begin{equation}\label{priceCDStheo}
\cds(t,T_a,T_b,R,\lgd) =
\mathbb{E}\{\pcds_{a,b}(t)|\mathcal{G}_t\} =
\mathbb{E}\{\pcds_{a,b}(t)|\mathcal{F}_t\} =:
\mathbb{E}_t\{\pcds_{a,b}(t)\}
\end{equation}
in our structural model setup. A CDS is quoted through its ``fair"
$R$, in that the rate $R$ that is quoted by the market at time $t$
satisfies $\cds(t,T_a,T_b,R,\lgd) = 0$. Let us assume, for
simplicity, deterministic interest rates; then we have
\begin{eqnarray}\label{priceCDS}
\nonumber \cds(t,T_a,T_b,R,\lgd) &:=&
-R\,\mathbb{E}_t\{P(t,\tau)(\tau-T_{\beta(\tau)-1})\mathbf{1}
_{\{T_a<\tau<T_b\}}\} \\
\nonumber &&-\sum_{i=a+1}^bP(t,T_i)
\alpha_iR\,\mathbb{E}_t\{\mathbf{1}_{\{\tau\geq T_i\}}\} +
\lgd\,\mathbb{E}_t\{\mathbf{1}_{\{T_a<\tau\leq T_b\}}P(t,\tau)\}\\
\nonumber &=&-R\left[\sum_{i=a+1}^b\left(P(t,T_i)
\alpha_i\,\mathbb{Q}\{\tau\geq T_i\}+\int_{T_{i-1}}^{T_i}(u-T_{i-1})P(t,u)d\mathbb{Q}(\tau\leq u)\right)\right]\\
&&+\lgd\int_{T_a}^{T_b}P(t,u)d\mathbb{Q}(\tau\leq u)
\end{eqnarray}

From our earlier
definitions, straightforward computations lead to the price at
initial time $0$ of a CDS, under deterministic interest rates, as
\begin{eqnarray}\label{cdsformulaat1p}
\cds_{a,b}(0,R,\lgd) = R \int_{T_a}^{T_b} P(0,t)
(t-T_{\beta(t)-1}) d \Qx(\tau > t)\\ \nonumber - R \sum_{i=a+1}^b
P(0,T_i) \alpha_i \Qx(\tau \ge T_i)
 - \lgd \int_{T_a}^{T_b} P(0,t) d \Qx(\tau
> t)
\end{eqnarray}
so that if one has a formula for the curve of survival
probabilities $t \mapsto \Qx(\tau
> t)$, as in our AT1P structural model, one also has a formula for CDS. It is clear that the fair rate $R$ strongly depends on the default
probabilities. The idea is to use quoted values of these fair
$R$'s with different maturities to derive the default
probabilities assessed by the market.

Formula~(\ref{survcoher}) can be used to fit the model parameters
to market data. If we aim at creating a one to one correspondence
between volatility parameters and CDS quotes, we can exogenously choose
the value $H$ and $B$, leaving all the unknown
information in the calibration of the volatility. If we do so, we
find exactly one volatility parameter for each CDS maturity,
including the first one. In our tests we have followed this approach,
where $H$ has been chosen externally before calibration. 
In \cite{BrTar04} and \cite{BrMorTar} we also suggest a methodology that takes into account equity volatilities in the calibration. 

In general the above CDS calibration procedures are justified by
the fact that in the end we are not interested in estimating the
real process of the firm value underlying the contract, but only
in reproducing risk neutral default probabilities with a model
that makes sense also economically. While it is important that the
underlying processes have an economic interpretation, we are not
interested in sharply estimating them or the capital structure of
the firm, but rather we appreciate the structural model
interpretation as a tool for assessing the realism of the outputs
of calibrations, and as an instrument to check economic
consequences and possible diagnostics.

In the following section we analyze how the AT1P model works in practice, in particular we consider the case of Lehman Brothers, one of the world major banks that incurred into a deep crisis, ending up with the bank's default. For simplicity our tests have been performed using the approximated postponed payoff for CDS (see \cite{BrMorTar} or Brigo and Mercurio (2006) for the details). An analysis of the same model on the Parmalat crisis, terminating in the 2003 default, is available in Brigo and Tarenghi (2004) and Brigo and Morini (2006).

\section{A Case Study with AT1P: Lehman Brothers default history}\label{sec:lehmanat1p}

\begin{itemize}
\item \textbf{August 23, 2007}: Lehman announces that it is going to shut one of its home lending units (\textit{BNC Mortgage}) and lay off 1,200 employees. The bank says it would take a \$52 million charge to third-quarter earnings.
\item \textbf{March 18, 2008}: Lehman announces better than expected first-quarter results (but profits have more than halved).
\item \textbf{June 9, 2008}: Lehman confirms the booking of a \$2.8 billion loss and announces plans to raise \$6 billion in fresh capital by selling stock. Lehman shares lose more than 9\% in afternoon trade.
\item \textbf{June 12, 2008}: Lehman shakes up its management; its chief operating officer and president, and its chief financial officer are removed from their posts.
\item \textbf{August 28, 2008}: Lehman prepares to lay off 1,500 people. The Lehman executives have been knocking on doors all over the world seeking a capital infusion.
\item \textbf{September 9, 2008}: Lehman shares fall 45\%.
\item \textbf{September 14, 2008}: Lehman files for bankruptcy protection and hurtles toward liquidation after it failed to find a buyer.
\end{itemize}


Here we show how the AT1P model calibration behaves when the credit quality of the considered company deteriorates in time (perceived as a widening of CDS spreads\footnote{It is market practice for CDS of names with deteriorating credit quality to quote an upfront premium rather than a running spread. After the so called ISDA Big Bang, it is likely that several names will quote an upfront on top of a fixed pre-specified running spread even when the credit quality has not deteriorated. In our tests we directly deal with the equivalent running spread alone, which can be obtained by upfront premia by means of standard techniques, see for example Beumee et al. (2009).}). We are going to analyze three different situations: i) a case of relatively stable situation, before the beginning of the crisis, ii) a case in the mid of the crisis and iii) a last case just before the default.

During our calibration we fix $\rec=40\%$, $B=0$ and $H=0.4$; this last choice is a completely arbitrary choice and has been suggested by the analogy with the CDS recovery rate. Also, as a comparison, we report the results of the calibration obtained using an intensity model. In simple intensity models, the survival probability can be computed as $\mathbb{Q}\left(\tau> t\right)=\exp\left(-\int_0^t\lambda(u)du\right)$, where $\lambda$ is the \textit{intensity function} or \textit{hazard rate} (assumed here to be deterministic). We choose a piecewise constant shape for $\lambda(t)$ and calibrate it to CDS quotes.

\subsection{Lehman Brothers CDS Calibration: July 10th, 2007}\label{sec:lehmanat1p_1}

On the left part of Table \ref{lehmanat1p_res1} we report the values of the quoted CDS spreads on July 10th, 2007, before the beginning of the crisis. We see that the spreads are very low, indicating a stable situation for Lehman. In the middle of Table \ref{lehmanat1p_res1} we have the results of the calibration obtained using an intensity model, while on the right part of the Table we have the results of the calibration obtained using the AT1P model.

It is important to stress the fact that the AT1P model is flexible enough to achieve exact calibration.

\begin{table}[h!]
\begin{center}
\begin{tabular}{|c|c||c|c||c|c|}
\hline
$T_i$ & $R_i$ (bps)& $\lambda_i$ (bps) & Surv (Int)& $\sigma_i$ & Surv (AT1P)\\
\hline
10 Jul 2007	&    &	     	 & 100.0\%	&        	& 100.0\%\\
1y					& 16 & 0.267\% &  99.7\%	&	29.2\%	&  99.7\%\\
3y					& 29 & 0.601\% &  98.5\%	&	14.0\%	&  98.5\%\\
5y					& 45 & 1.217\% &  96.2\%	&	14.5\%	&  96.1\%\\
7y					& 50 & 1.096\% &  94.1\%	&	12.0\%	&  94.1\%\\
10y					& 58 & 1.407\% &  90.2\%	&	12.7\%	&  90.2\%\\
\hline
\end{tabular}
\end{center}
\caption{\small Results of calibration for July 10th, 2007.}\label{lehmanat1p_res1}
\end{table}

\subsection{Lehman Brothers CDS Calibration: June 12th, 2008}\label{sec:lehmanat1p_2}

In Table \ref{lehmanat1p_res2} we report the results of the calibration on June 12th, 2008, in the middle of the crisis. We see that the CDS spreads $R_i$ have increased with respect to the previous case, but are not very high, indicating the fact that the market is aware of the difficulties suffered by Lehman but thinks that it can come out of the crisis.
The survival probability resulting from calibration is lower than in the previous case; since the barrier parameter $H$ has not changed, this translates into higher volatilities.

\begin{table}[h!]
\begin{center}
\begin{tabular}{|c|c||c|c||c|c|}
\hline
$T_i$ & $R_i$ (bps) & $\lambda_i$ (bps) & Surv (Int)& $\sigma_i$ & Surv (AT1P)\\
\hline
12 Jun 2008	&     &    			& 100.0\%	&        	& 100.0\%\\
1y					& 397 & 6.563\% &  93.6\%	&	45.0\%	&  93.5\%\\
3y					& 315 & 4.440\% &  85.7\%	&	21.9\%	&  85.6\%\\
5y					& 277 & 3.411\% &  80.0\%	&	18.6\%	&  79.9\%\\
7y					& 258 & 3.207\% &  75.1\%	&	18.1\%	&  75.0\%\\
10y					& 240 & 2.907\% &  68.8\%	&	17.5\%	&  68.7\%\\
\hline
\end{tabular}
\end{center}
\caption{\small Results of calibration for June 12th, 2008.}\label{lehmanat1p_res2}
\end{table}

\subsection{Lehman Brothers CDS Calibration: September 12th, 2008}\label{sec:lehmanat1p_3}

In Table \ref{lehmanat1p_res3} we report the results of the calibration on September 12th, 2008, just before Lehman's default. We see that the spreads are now very high, corresponding to lower survival probability and higher volatilities than before.
\begin{table}[h!]
\begin{center}
\begin{tabular}{|c|c||c|c||c|c|}
\hline
$T_i$ & $R_i$ (bps) & $\lambda_i$ (bps) & Surv (Int)& $\sigma_i$ & Surv (AT1P)\\
\hline
12 Sep 2008	&    	 &          & 100.0\%	&        	& 100.0\%\\
1y					& 1437 & 23.260\% &  79.2\%	&	62.2\%	&  78.4\%\\
3y					&  902 & 9.248\%  &  65.9\%	&	30.8\%	&  65.5\%\\
5y					&  710 & 5.245\%  &  59.3\%	&	24.3\%	&  59.1\%\\
7y					&  636 & 5.947\%  &  52.7\%	&	26.9\%	&  52.5\%\\
10y					&  588 & 6.422\%  &  43.4\%	&	29.5\%	&  43.4\%\\
\hline
\end{tabular}
\end{center}
\caption{\small Results of calibration for September 12th, 2008.}\label{lehmanat1p_res3}
\end{table}

\subsection{Comments}\label{sec:at1p_lehm_comments}
We have seen that the AT1P model can calibrate exactly CDS market quotes, and the survival probabilities obtained are in accordance with those obtained using an intensity model. This confirms the well known fact that, when interest rates are assumed independent of default, survival probabilities can be implied from CDS in a model-independent way (formula (\ref{cdsformulaat1p})). Anyway, after a deeper analysis of the results, we find:
\begin{itemize}
\item[-] \textit{Scarce relevance of the barrier in calibration}: the barrier parameter $H$ has been fixed before calibration and everything is left to the volatility calibration;
\item[-] \textit{High discrepancy between first volatility bucket and the following values}.
\end{itemize}
The problem is that when the default boundary is deterministic, diffusion models tend to calibrate a relevant probability of default by one-year (shortest horizon credit spread) only by supposing particularly high one-year volatility.  This is due to the fact that, initially, with low volatilities the trajectories of a model like (\ref{valueproc}) do not widen fast enough to hit a deterministic barrier frequently enough to generate relevant default probabilities. One has therefore to choose a high initial volatility to achieve this. The problem is also related to the fundamental assumption that the default threshold is a deterministic, known function of time, based on reliable accounting data. This is a very strong assumption and usually it is not true: balance sheet information is not certain, possibly because the company is hiding information, or because a real valuation of the firm assets is not easy (for example in case of derivative instruments). Public investors, then, may have only a partial and coarse information about the true value of the firm assets or the related liability-dependent firm condition that would trigger default.

In AT1P model, $H$ is the ratio between the initial level of the default barrier and the initial value of the company assets. To take market uncertainty into account in a realistic albeit simple manner, $H$ can be replaced by a random variable assuming different values in different scenarios.
This is the main idea which leads us to introduce the SBTV model.

\section{SBTV Model (Brigo and Morini, 2006)}\label{sec:sbtv}

Where may one consider explicitly market uncertainty on the situation of the company,
due to the fact that balance sheet information is not always reliable or easy to value? This is a possibility  in case the company is hiding information, or in case illiquidity causes the valuation of the firm assets and liabilities to be uncertain. In the first case, following for
example Giesecke (2004) who refers to scandals such as Enron, Tyco and WorldCom, a
crucial aspect in market uncertainty is that public investors have only a partial and coarse
information about the true value of the firm assets or the related liability-dependent firm
condition that would trigger default.  This was our motivation to introduce randomness in $H$ when we dealt with Parmalat in \cite{BrTar05} and \cite{BrMor}. In particular, randomness in the initial level $H$ of the barrier was used in \cite{BrMor} to represent the uncertainty of market operators on the real financial situation of Parmalat, due to lack of transparency and actual fraud in Parmalat's accounting. Here it can represent equally well the uncertainty of market operators on the real financial situation of Lehman. But in this case the uncertainty is related to an objective difficulty in assigning a fair value to a large part of the assets and liabilities of Lehman (illiquid mortgage-related portfolio credit derivatives) and to the intrinsic complexity of the links between the bank and the related SIVs and conduits.

Therefore, in order to take market uncertainty into account
in a realistic, albeit simple, manner, in the following $H$ is replaced by a random variable
assuming different values in different scenarios, each scenario with a different probability.

As in Brigo and Morini (2006)\cite{BrMor}, in this work we deem scenarios on the barrier to
be an efficient representation of the uncertainty on the balance sheet of a company, while
deterministic time-varying volatility can be required for precise and efficient calibration
of CDS quotes. The resulting model is called Scenario Barrier Time-Varying Volatility AT1P Model (SBTV). In this way we can achieve exact calibration to all market quotes.
Let the assets value $V$ risk neutral dynamics be given by (\ref{valueproc}). The default time $\tau$ is
again the first time where $V$ hits the barrier from above, but now we have a scenario
barrier
\begin{equation*}
H^I(t)=H^I\exp\left(\int_0^t(r_u-k_u-B\sigma^2_u)du\right)=\frac{H^I}{V_0}\mathbb{E}\left[V_t\right]\exp\left(-B\int_0^t\sigma^2_u du\right)
\end{equation*}
where $H^I$ assumes scenarios $H^1$, $H^2$,..., $H^N$ with $\mathbb{Q}$ probabilities $p^1$, $p^2$,..., $p^N$. All probabilities are in $[0,1]$ and add up to one, and $H^I$ is independent of $W$. Thus now the ratio $H^I/V_0$ depends on the scenario $I$. If we are to price a default-sensitive discounted payoff $\Pi$, by iterated expectation we have
\begin{equation*}
\mathbb{E}\left[\Pi\right]=\mathbb{E}\left[\mathbb{E}\left[\Pi|H^I\right]\right]=\sum_{i=1}^Np^i\,\mathbb{E}\left[\Pi|H^I=H^i\right]
\end{equation*}
so that the price of a security is a weighted average of the prices of the security in the
different scenarios, with weights equal to the probabilities of the different scenarios. For
CDS, the price with the SBTV model is
\begin{equation}\label{sbtv_cds}
\mbox{SBTV}\textbf{\cds}_{a,b}=\sum_{i=1}^N p^i \cdot \mbox{AT1P}\textbf{\cds}_{a,b}(H^i)
\end{equation}
where $AT1P\textbf{\cds}_{a,b}(H^i)$ is the CDS price computed according to the AT1P survival probability formula (\ref{survcoher}) when $H=H^i$. Hence, the SBTV model acts like a mixture of AT1P scenarios.

\section{A Case Study with SBTV: Lehman Brothers default history}\label{sec:lehmansbtv}

Here we want to show how the SBTV model calibration behaves with respect to the AT1P model. We consider the Lehman Brothers example as before. We limit our analysis to only two barrier scenarios ($H^1$ with probability $p^1$ and $H^2$ with probability $p^2=1-p^1$), since, according to our experience, further scenarios do not increase the calibration power.

In the calibration, we set the lower barrier parameter $H^1=0.4$. If we consider $M$ CDS quotes, we have $M+2$ unknown parameters: $H^2$, $p^1$ ($p^2=1-p^1$) and all the volatilities $\sigma_j$ corresponding to the $M$ buckets. It is clear that a direct fit like in the AT1P case (where we have one unknown volatility $\sigma_j$ for each CDS quote $R_j$) is not possible. Exact calibration could be achieved using a two steps approach:
\begin{itemize}
\item[1]: We limit our attention to the first three CDS quotes, and set $\sigma_1=\sigma_2=\sigma_3=\bar{\sigma}$. Now we have three quotes for three unknowns ($H^2$, $p^1$ and $\bar{\sigma}$) and can run a best-fit on these parameters (not exact calibration, since the model could not be flexible enough to attain it).
\item[2]: At this point we go back to consider all the $M$ CDS quotes and, using $H^2$ and the $p$'s just obtained, we can run a second step calibration on the $M$ volatilities $\sigma_j$ to get an exact fit. Notice that if the first step calibration is good enough, then the refinement to $\sigma_{1,2,3}$ due to second step calibration is negligible.
\end{itemize}

\subsection{Lehman Brothers CDS Calibration: July 10th, 2007}\label{sec:lehmansbtv_1}

In Table \ref{lehm_scen1} we report the values of the calibrated barrier parameters with their corresponding probabilities, while in Table \ref{lehmansbtv_res1} we show the results of the calibration.
\begin{table}[h!]
\begin{center}
\begin{tabular}{|c|c|c|}
\hline
$\mbox{Scenario}$ & $H$ & $p$\\
\hline
1 & 0.4000 & 96.2\%\\
2 & 0.7313 &  3.8\%\\
\hline
\end{tabular}
\end{center}
\caption{\small Scenario barriers and probabilities.}\label{lehm_scen1}
\end{table}

\begin{table}[h!]
\begin{center}
\begin{tabular}{|c|c||c|c||c|c|}
\hline
$T_j$ & $R_j$ (bps) & $\sigma_j$ (bps) & Surv (SBTV)& $\sigma_j$ & Surv (AT1P)\\
\hline
10 Jul 2007	&    &      & 100.0\%	&        	& 100.0\%\\
1y					& 16 & 16.6\% &  99.7\%	&	29.2\%	&  99.7\%\\
3y					& 29 & 16.6\% &  98.5\%	&	14.0\%	&  98.5\%\\
5y					& 45 & 16.6\% &  96.1\%	&	14.5\%	&  96.1\%\\
7y					& 50 & 12.6\% &  94.1\%	&	12.0\%	&  94.1\%\\
10y					& 58 & 12.9\% &  90.2\%	&	12.7\%	&  90.2\%\\
\hline
\end{tabular}
\end{center}
\caption{\small Results of calibration for July 10th, 2007.}\label{lehmansbtv_res1}
\end{table}
Looking at the results in Tables \ref{lehm_scen1} and \ref{lehmansbtv_res1} we see that, in the case of the quite stable situation for Lehman, we have a lower barrier scenario (better credit quality) with very high probability, and a higher barrier scenario (lower credit quality)  with low probability. Also, when comparing the results with the AT1P calibration, we see that now the calibrated volatility is nearly constant on all maturity buckets, which is a desirable feature for the firm value dynamics.

\subsection{Lehman Brothers CDS Calibration: June 12th, 2008}\label{sec:lehmansbtv_2}

\begin{table}[h!]
\begin{center}
\begin{tabular}{|c|c|c|}
\hline
$\mbox{Scenario}$ & $H$ & $p$\\
\hline
1 & 0.4000 & 74.6\%\\
2 & 0.7971 & 25.4\%\\
\hline
\end{tabular}
\end{center}
\caption{\small Scenario barriers and probabilities.}\label{lehm_scen2}
\end{table}

\begin{table}[h!]
\begin{center}
\begin{tabular}{|c|c||c|c||c|c|}
\hline
$T_j$ & $R_j$ (bps) & $\sigma_j$ (bps) & Surv (SBTV)& $\sigma_j$ & Surv (AT1P)\\
\hline
12 Jun 2008	&     &        & 100.0\%	&        	& 100.0\%\\
1y					& 397 & 18.7\% &  93.6\%	&	45.0\%	&  93.5\%\\
3y					& 315 & 18.7\% &  85.7\%	&	21.9\%	&  85.6\%\\
5y					& 277 & 18.7\% &  80.1\%	&	18.6\%	&  79.9\%\\
7y					& 258 & 17.4\% &  75.1\%	&	18.1\%	&  75.0\%\\
10y					& 240 & 16.4\% &  68.8\%	&	17.5\%	&  68.7\%\\
\hline
\end{tabular}
\end{center}
\caption{\small Results of calibration for June 12th, 2008.}\label{lehmansbtv_res2}
\end{table}
Looking at the results in Tables \ref{lehm_scen2} and \ref{lehmansbtv_res2} we see that the (worse credit quality) barrier parameter $H^2$ has both a higher value (higher proximity to default) and a much higher probability with respect to the calibration case of July 10th, 2007. This is due to the higher CDS spread values. Moreover, by noticing that the fitted volatility has not increased too much, we can argue that the worsened credit quality can be reflected into a higher probability of being in the scenario with higher default barrier (worsened credit quality).

\subsection{Lehman Brothers CDS Calibration: September 12th, 2008}\label{sec:lehmansbtv_3}

Here we have a large increase in CDS spreads, which can be explained by a very large probability of $50\%$ for the higher barrier scenario and a higher value for the scenario itself, that moves to 0.84 from the preceding cases of 0.79 and 0.73. Tables \ref{lehm_scen3} and \ref{lehmansbtv_res3} show the results of calibration. We see that there are not greater differences in volatilities than before, and the larger default probability can be explained by a higher level of proximity to default and high probability of being in that proximity  (high $H$ scenario). In this particular case we have equal probability of being either in the risky or in the stable scenario.

\begin{table}[h!]
\begin{center}
\begin{tabular}{|c|c|c|}
\hline
$\mbox{Scenario}$ & $H$ & $p$\\
\hline
1 & 0.4000 & 50.0\%\\
2 & 0.8427 & 50.0\%\\
\hline
\end{tabular}
\end{center}
\caption{\small Scenario barriers and probabilities.}\label{lehm_scen3}
\end{table}

\begin{table}[h!]
\begin{center}
\begin{tabular}{|c|c||c|c||c|c|}
\hline
$T_j$ & $R_j$ (bps) & $\sigma_j$ (bps) & Surv (SBTV)& $\sigma_j$ & Surv (AT1P)\\
\hline
12 Sep 2008	&      &        & 100.0\%	&        	& 100.0\%\\
1y					& 1437 & 19.6\% &  79.3\%	&	62.2\%	&  78.4\%\\
3y					&  902 & 19.6\% &  66.2\%	&	30.8\%	&  65.5\%\\
5y					&  710 & 19.6\% &  59.6\%	&	24.3\%	&  59.1\%\\
7y					&  636 & 21.8\% &  52.9\%	&	26.9\%	&  52.5\%\\
10y					&  588 & 23.7\% &  43.6\%	&	29.5\%	&  43.4\%\\
\hline
\end{tabular}
\end{center}
\caption{\small Results of calibration for September 12th, 2008.}\label{lehmansbtv_res3}
\end{table}

\subsection{Comments}\label{sec:sbtv_lehm_comments}

We have seen that the calibration performed with the SBTV model is comparable, in terms of survival probabilities, with the calibration obtained using the AT1P model. Also, the calibration is exact in both cases. However, we have seen that the SBTV model returns a more stable volatility term structure, and also has a more robust economic interpretation.

One of the main drawbacks of structural models is that they usually are not able to explain short term credit spreads; in fact, usually the diffusion part of the GBM is not enough to explain a non null default probability in very small time intervals. The introduction of default barrier scenarios could be a way to overcome this problem; in fact, a very high scenario barrier could be enough to account for short term default probabilities.

At this point, a natural extension of the family of structural models we have presented here is the valuation of hybrid equity/credit products. In the following part of this paper, we are going to deal with this application.

\section{A fundamental example: Pricing Counterparty Risk in Equity Return Swaps}\label{sec:otherproducts}
In this section we present an example of pricing with the
calibrated structural model. This follows Brigo and Tarenghi (2004), and is reported also in the last part of Brigo and Masetti (2005). This example concerns the valuation
of an equity return swap where we take into account counterparty risk,
and is chosen to highlight one case where the calibrated
structural model may be preferable to a reduced-form intensity
model calibrated to the same market information. This is an
illustration of a more general situation that typically occurs
when one tries to price the counterparty risk in an equity payoff.
We will see that it is possible to split the expectation of the
payoff, and that the decomposition roughly involves the valuation
of the same payoff without counterparty risk and the valuation of
an option on the residual NPV of the considered payoff at the
default time of the counterparty. Therefore including the
counterparty risk adds an optionality level to the payoff.

Let us consider an equity return swap payoff. Assume we are a company
``A" entering a contract with company ``B", our counterparty. The
reference underlying equity is company ``C". We assume ``A'' to be default-free, i.e. we consider unilateral counterparty risk. For a discussion on unilateral vs bilateral risk, see Brigo and Capponi (2008) and Brigo, Pallavicini and Papathodorou (2009). The contract, in its
prototypical form, is built as follows. Companies ``A" and ``B"
agree on a certain amount $K$ of stocks of a reference entity ``C"
(with price $S$) to be taken as nominal ($N=K\,S_0$). The contract
starts in $T_a=0$ and has final maturity $T_b = T$. At $t=0$ there
is no exchange of cash (alternatively, we can think that ``B"
delivers to ``A" an amount $K$ of ``C" stock and receives a cash
amount equal to $K S_0$). At intermediate times ``A" pays to ``B"
the dividend flows of the stocks (if any) in exchange for a
periodic rate (for example a semi-annual LIBOR or EURIBOR rate
$L$) plus a spread $X$. At final maturity $T=T_b$, ``A" pays $K
S_T$ to ``B" (or gives back the amount $K$ of stocks) and receives
a payment $K S_0$. This can be summarized as in (\ref{for:ERScashflows}).

\begin{eqnarray}\label{for:ERScashflows}
 \mbox{Initial Time 0:} &\mbox{NO FLOWS, or}&   \\
{\Huge{\boxed{A}}} \ \  \longrightarrow &K S_0\ \mbox{cash}&
\longrightarrow  \ \ \ \ {\Huge{\boxed{B}}}  \nonumber \\
   \longleftarrow  &K\ \mbox{equity}& \longleftarrow  \nonumber \\
&....& \nonumber  \\
\mbox{Time}\ T_i:\ \  \longrightarrow &\mbox{equity dividends}& \longrightarrow  \nonumber \\
{\Huge{\boxed{A}}} \ \  \longleftarrow  &\mbox{Libor + Spread}& \longleftarrow  \ \ \ \ {\Huge{\boxed{B}}} \nonumber \\
&....&  \nonumber \\
 \mbox{Final Time}\ T_b:\ \  \longrightarrow &K\ \mbox{equity} \ \mbox{or} \ K S_{T_b}\ \mbox{cash}&
\longrightarrow
 \nonumber \\
{\Huge{\boxed{A}}} \ \ \longleftarrow  &K S_0\ \mbox{cash}& \longleftarrow   \ \ \ \ {\Huge{\boxed{B}}} \nonumber
\end{eqnarray}

The price of this product can be derived using risk neutral
valuation, and the (fair) spread is chosen in order to obtain a
contract whose value at inception is zero. We ignore default of
the underlying ``C", thus assuming it has a much stronger credit
quality than the counterparty ``B". This can be the case for
example when ``C" is an equity index (Pignatelli~(2004)). It can
be proved that if we do not consider default risk for ``B", the
fair spread is identically equal to zero. But when taking into
account counterparty default risk in the valuation the fair spread
is no longer zero.   In case an early default of the counterparty
``B" occurs, the following happens.  Let us call $\tau=\tau_B$ the
default instant. Before $\tau$ everything is as before,  but if
$\tau\leq T$, the net present value (NPV) of the position at time
$\tau$ is computed. If this NPV is negative for us, i.e. for ``A",
then its opposite is completely paid to ``B" by us at time $\tau$
itself. On the contrary, if it is positive for ``A", it is not
received completely but only a recovery fraction $\rec$ of that
NPV is received by us. It is clear that to us (``A") the
counterparty risk is a problem when the NPV is large and positive,
since in case ``B" defaults we receive only a fraction of it.

Analytically, the risk neutral expectation of the discounted
payoff is ($L(S,T)$ is the simply compounded rate at time $S$ for
maturity $T$):
\begin{eqnarray}\label{ESpayoff}
\Pi_{ES}(0) & = &
\mathbb{E}_0\bigg\{\mathbf{1}_{\{\tau>T_b\}}\bigg[-K\,\npv_{dividends}^{0-T_b}(0)+ KS_0\,\sum_{i=1}^b D(0,T_i)\alpha_i\big(L(T_{i-1},T_i)
\nonumber\\
& & +X\big)+D(0,T_b)\big(KS_0-KS_{T_b}\big)\bigg] \nonumber\\
& &  +\mathbf{1}_{\{\tau\leq
T_b\}}\bigg[-K\,\npv_{dividends}^{0-\tau}(0)+KS_0\,\sum_{i=1}^{\beta(\tau)-1}D(0,T_i)\alpha_i\big(L(T_{i-1},T_i)
\nonumber\\
& & +X\big)+D(0,\tau)\big(\rec(\npv(\tau))^+
-(-\npv(\tau))^+\big)\bigg]\bigg\}
\end{eqnarray}
where
\begin{eqnarray}\label{NPVdef}
\npv(\tau) & = &
\mathbb{E}_{\tau}\bigg\{-K\,\npv_{dividends}^{\tau-T_b}(\tau)+KS_0\,\sum_{i=\beta(\tau)}^{b}
D(\tau,T_i)\alpha_i\left(L(T_{i-1},T_i)+X\right) \nonumber \\
& & +\left(KS_0-KS_{T_b}\right)D\left(\tau,T_b\right)\bigg\}.
\end{eqnarray}
We denote by $\npv_{dividends}^{s-t}(u)$  the net present
value of the dividend flows between $s$ and $t$ computed in $u$.

We state the following\footnote{A proof of the proposition is given in the most general setup in Brigo and Masetti (2005).}:
\begin{proposition}{\bf (Equity Return Swap price under
Counterparty Risk)}. The fair price of the Equity Swap defined
above, i.e.~(\ref{ESpayoff}), can be simplified as follows:
\begin{eqnarray}\label{ESprop}
\Pi_{ES}(0) = KS_0 X \sum_{i=1}^b\alpha_i
P(0,T_i)-\lgd\,\mathbb{E}_0\bigg\{ \mathbf{1}_{\{\tau\leq
T_b\}}P(0,\tau)(\npv(\tau))^+\bigg\}.
\end{eqnarray}
The first term is the equity swap price in a default-free world,
whereas the second one is the optional price component due to
counterparty risk.
\end{proposition}

If we try and find the above price by computing the expectation
through a Monte Carlo simulation, we have to simulate both the
behavior of the equity ``C" underlying the swap, which we call
$S_t=S_t^C$,  and the default of the counterparty ``B". In
particular we need to know exactly $\tau=\tau_B$. Obviously the
correlation between ``B" and ``C" could have a relevant impact on
the contract value. Here the structural model can be helpful:
Suppose to calibrate the underlying process $V$ to CDS's for name
``B", finding the appropriate default barrier and volatilities
according to the procedure outlined earlier in this paper with the
AT1P or SBTV model. We could set a correlation between the processes
$V^B_t$ for ``B" and $S_t$ for ``C", derived for example through
historical estimation directly based on equity returns, and
simulate the joint evolution of $[V^B_t, S_t]$. As a proxy of the
correlation between these two quantities we may consider the
correlation between $S^B_t$ and $S^C_t$, i.e. between equities.

Going back to our equity swap, now it is possible to run the Monte
Carlo simulation, looking for the spread $X$ that makes the
contract fair. The simulation itself is simpler when taking into
account the following computation included in the discounted NPV:
\begin{equation}\label{exptaufinalprice}
P(\tau,T_b)\mathbb{E}_{\tau}\big\{S_{T_b}\big\} = S_{\tau}
-\npv_{dividends}^{\tau-T_b}(\tau)
\end{equation}
so that we have
\begin{eqnarray}\label{NPVsimpl2}
\nonumber P(0,\tau) \npv(\tau) & = &
KS_0\,\sum_{i=\beta(\tau)}^{b} P(0,T_i)\alpha_i(L(T_{i-1},T_i)+X)+KS_0 P(0,T_b)\\
& & \nonumber - K P(0,\tau) S_{\tau} =\\
\nonumber & = & KS_0\,\sum_{i=\beta(\tau)}^{b} P(0,T_i)\alpha_i X
+ KS_0 P(0,T_{\beta(\tau)-1})\\ & &  - K P(0,\tau) S_{\tau}.
\end{eqnarray}

The reformulation of the original expected payoff (\ref{ESpayoff})
as in (\ref{ESprop}) presents an important advantage in terms of
numerical simulation. In fact in (\ref{ESpayoff}) we have a global
expectation, hence we have to simulate the exact payoff for each
path. In (\ref{ESprop}), with many simplifications, we have
isolated a part of the expected payoff out of the main
expectation. This isolated part has an expected value that we have
been able to calculate, so that it does not have to be simulated.
Simulating only the residual part is helpful because now the
variance of the part of the payoff that has been computed
analytically is no longer affecting the standard error of our
Monte Carlo simulation. The standard error is indeed much lower
when simulating (\ref{ESprop}) instead of (\ref{ESpayoff}). The
expected value we computed analytically above involves terms in
$S_{T_b}$ which would add a lot of variance to the final payoff.
In (\ref{ESprop}) the only $S_{T_b}$ term left is in the optional
NPV part.

We performed some simulations under different assumptions on the
correlation between ``B" and ``C". We considered five cases:
$\rho= -1$, $\rho = -0.2$, $\rho = 0$, $\rho = 0.5$ and $\rho =
1$. In Table \ref{EScorrelation} we present the results of the
simulation. For counterparty ``B" we
used the CDS rates reported in Table \ref{cds_es}. For the reference stock ``C" we used a
hypothetical stock with initial price $S_0=20$, volatility $\sigma
= 20\%$ and constant dividend yield $q=0.80\%$. The simulation date is September 16th, 2009. The contract has
maturity $T=5y$ and the settlement of the LIBOR rate has a
semi-annual frequency. Finally, we included a recovery rate
$\rec=40\%$. Since the reference number of
stocks $K$ is just a constant multiplying the whole payoff,
without losing generality we set it equal to one.

\begin{table}
\begin{center}
\begin{tabular}{|c|c|c|}
\hline
$T_i$ & $R^{BID}_i$ (bps) & $R^{ASK}_i$ (bps) \\
\hline
1y  &  25  &  31\\
3y  &  34  &  39\\
5y  &  42  &  47\\
7y  &  46  &  51\\
10y &  50  &  55\\
\hline
\end{tabular}
\end{center}
\caption{\small CDS spreads used for the counterparty ``B" credit quality in the valuation of the equity return swap.}\label{cds_es}
\end{table}

In order to reduce the errors of the simulations, we adopted
a variance reduction technique using the default indicator (whose
expected value is the known default probability) as a control
variate. In particular we have used the default indicator
$1_{\{\tau < T\}}$ at the maturity $T$ of the contract, which has
a large correlation with the final payoff. Even so, a large number
of scenarios are needed to obtain errors with a lower order of
magnitude than $X$.

We notice that $X$ increases together with $\rho$. This fact can
be explained in the following way. Let us consider the case of
positive correlation between ``B" and ``C": This means that, in
general, if the firm value for ``B" increases, moving away from
the default barrier, also the stock price for ``C" tends to
increase due to the positive correlation. Both processes will then
have high values. Instead, again under positive correlation, if
$V^B_t$ lowers towards the default barrier, also $S^C_t$ will tend
to do so, going possibly below the initial value $S_0$. In this
case $NPV(\tau)$ has a large probability to be positive (see
(\ref{NPVsimpl2})), so that one needs a large $X$ to balance it,
as is clear when looking at the final payoff (\ref{ESprop}). On
the contrary, for negative correlation, the same reasoning can be
applied, but now if $V^B_t$ lowers and tends to the  default
barrier, in general $S^C_t$ will tend to move in the opposite
direction and the corresponding $NPV(\tau)$ will probably be
negative, or, if positive, not very large. Hence the ``balancing"
spread $X$ we need will be quite small.

\begin{table}[h!]
\begin{center}
\begin{tabular}{|c|c|c|}
\hline $\rho$ & X (AT1P) & X (SBTV) \\
\hline
-1      &   0.0     &   0.0     \\
-0.2    &   3.0     &   3.6     \\
0       &   5.5     &   5.5     \\
0.5     &   14.7    &   11.4    \\
1       &   24.9    &   17.9    \\
\hline
\end{tabular}
\end{center}
\caption{\small Fair spread $X$ (in basis points) of the Equity Return
Swap in five different correlation cases for AT1P and SBTV models.}\label{EScorrelation}
\end{table}

Also we notice that, for $\rho\neq 0$, the fair spread $X$ is different when computed either with the AT1P or the SBTV model. This can be explained by the fact that the payoff depends strongly on the dependence structure between $\tau$ and $S_T$, while $\rho$ is an instantaneous correlation that ignores randomness in the barrier that defines $\tau$; because of the different volatility term structure obtained with the two models, but especially because of the random barrier in the SBTV model, we have that the same value of $\rho$ corresponds to different dependence structures between $\tau$ and $S_T$, and hence to different values for the fair spread $X$. This shows that counterparty risk pricing is quite subject to ``model risk", since two models calibrated to the same data give different answers.

Finally, as a comparison, we computed $X$ also with a calibrated intensity model, and obtained $X=5.5$ bps, a value consistent with the case $\rho=0$, representing the independence case.

\section{Conclusions}
In general the link between default probabilities and credit
spreads is best described by intensity models. The credit spread
to be added to the risk free rate represents a good measure of a
bond credit risk for example. Yet, intensity models present some
drawbacks: They do not link the default event to the economy but
rather to an exogenous jump process whose jump component remains
unexplained from an economic point of view. 

In this paper we introduced two analytically tractable structural models (AT1P and SBTV) that allow for a solution to the above points. In these
models the default has an economic cause, in that it is caused by
the value of the firm hitting the safety barrier value, and all
quantities are basic market observables. The extension to hybrid equity/credit products turns out to be natural, given the possibility to model dependency between equity assets and firms asset just by modelling the correlation between the two dynamics.

We showed how to calibrate the AT1P and SBTV model parameters to actual market
data: Starting from CDS quotes, we calibrated the value of the
firm volatilities that are consistent with CDS quotes and also leading to analytical
tractability. We also explained the analogies with barrier option
pricing, in particular the case with time dependent parameters.

As a practical example, we also applied the model to a concrete
case, showing how it can describe the proximity of default when
time changes and the market quotes for the CDS's reflect
increasing credit deterioration.  When the market detects a
company crisis, it responds with high CDS quotes and this
translates into high default probabilities, i.e. high
probabilities for the underlying process to hit the safety
barrier, that in turn translate in high calibrated volatilities
for the firm value dynamics.

Also, we showed how these two models can be used in practice: we analyzed the case of an equity return swap, evaluating the cost embedded in the instrument due to counterparty risk.


\begin{thebibliography}{50}

\bibitem{Alavian} Alavian, S., Ding, J., Whitehead, P., Laudicina, L. (2009) Counterparty Valuation Adjustment (CVA). Working paper, available at {\tt defaultrisk.com}.

\bibitem{Assefa} Assefa S., Bielecki T., Crepey S. and Jeanblanc M. (2009): {\em CVA computation for counterparty risk assessment in credit portfolio.} Preprint.

\bibitem{Beumee} Beumee J., Brigo D., Schiemert D. and Stoyle G. (2009): Charting a Course Through the CDS Big Bang. Fitch Solution research report.

\bibitem{BielRut} Bielecki T. and Rutkowski M. (2001):
\emph{Credit risk: Modeling, Valuation and Hedging.} Springer
Verlag.

\bibitem{blackcox} Black F. and Cox J. C.
(1976): Valuing corporate securities: Some effects of bond
indenture provisions. {\em J. of Finance} 31, 351-367.

\bibitem{Blanchet} Blanchet-Scalliet C. and Patras F. (2008) Counterparty Risk Valuation for CDS. Available at {\tt defaultrisk.com}.

\bibitem{Brigo04solo}
Brigo D. (2004):  Candidate Market Models and the Calibrated
CIR++ Stochastic Intensity Model for Credit Default Swap Options
and Callable Floaters. In {\em Proceedings of the 4-th ICS
Conference}, Tokyo, March 18-19, 2004. Available at
www.damianobrigo.it. Short version in ``Market Models for CDS
options and callable floaters", {\em Risk Magazine}, January 2005.

\bibitem{brigo04b} Brigo D. (2004b): Constant Maturity Credit
Default Swap Pricing with Market Models. Available at {\tt ssrn.com}. Short version in {\em Risk Magazine}, June 2006 issue.

\bibitem{BrBakk} Brigo D. and Bakkar I. (2009): Accurate counterparty risk valuation for energy-commodities swaps. {\em Energy Risk}, March issue.

\bibitem{BrCapp} Brigo D. and Capponi A. (2008): Bilateral Counterparty Risk Valuation with Stochastic Dynamical Models and Application to Credit Default Swaps. Working Paper. Available at {\tt ssrn.com} and {\tt arXiv.org}.

\bibitem{BrChour} Brigo D. and Chourdakis K. (2009): Counterparty Risk for Credit Default Swaps: Impact of spread volatility and default correlation. Accepted for publication in {\em International Journal of Theoretical and Applied Finance}.

\bibitem{BrEl} Brigo, D. and  El--Bachir, N. (2008). An exact formula for default swaptions pricing in the SSRJD stochastic intensity model. To appear in {\em Mathematical Finance.} Available at {\tt defaultrisk.com}, {\tt ssrn.com} and {\tt arXiv.org}.

\bibitem{Masetti} Brigo D. and Masetti M. (2005): Risk Neutral Pricing of Counterparty Risk. In: Pykhtin M. (editor) {\em Counterpaty Credit Risk Modeling: Risk Management, Pricing and Regulation}. Risk Books, London.

\bibitem{BrMer} Brigo D. and Mercurio F. (2006):
\emph{Interest Rate Models: Theory and Practice. With Smile, Inflation and Credit, 2nd edition.} Springer Verlag.

\bibitem{BrMor} Brigo D. and Morini M. (2006): Structural Credit Calibration, \textit{Risk Magazine}, April issue.

\bibitem{BrMorTar} Brigo D. and Morini M. and Tarenghi, M. (2010). Credit Calibration with Structural Models and Equity Return Swap valuation under Counterparty Risk, in: Bielecki, Brigo and Patras (Editors), Recent advancements in theory and practice of credit derivatives, Bloomberg Press, 2010, Forthcoming

\bibitem{BrPall} Brigo D. and Pallavicini A. (2007): Counterparty Risk under Correlation between Default and Interest Rates. In: Miller J., Edelman D. and Appleby J. (editors) {\em Numerical Methods for Finance}, Chapman Hall.

\bibitem{BrPallPap} Brigo D., Pallavicini A. and Papatheodorou V. (2009): Bilateral counterparty risk valuation for interest-rate products: impact of volatilities and correlations. Working Paper. Available at {\tt ssrn.com} and {\tt arXiv.org}.

\bibitem{BrTar04} Brigo D. and Tarenghi, M. (2004): Credit Default Swap Calibration and Equity Swap Valuation under Counterparty Risk with a Tractable Structural Model. Proceedings of the FEA 2004 Conference at MIT, Cambridge, Massachussets, November 8-10. Available at {\tt ssrn.com}, {\tt arXiv.org} and {\tt defaultrisk.com}.

\bibitem{BrTar05} Brigo D. and Tarenghi M. (2005): Credit Default Swap Calibration and Counterparty Risk Valuation with a Scenario based First Passage Model. Available at {\tt ssrn.com}, {\tt arXiv.org} and {\tt defaultrisk.com}.

\bibitem{Briys} Briys, E. and de Varenne, F. (1997). Valuing risky fixed rate debt: An extension. {\em Journal of Financial and Quantitative Analysis}, 32(2):239–-248.

\bibitem{CollDufr01} Collin-Dufresne P. and Goldstein R (2001): Do credit spreads reflect stationary leverage ratios?, {\em Journal of Finance} 56, 1929-1958.

\bibitem{Crepey} Crepey S., Jeanblanc M. and Zargari B. (2009): CDS with Counterparty Risk in a Markov Chain Copula Model with Joint Defaults. Working Paper.

\bibitem{crouhy00} Crouhy M., Galai D. and Mark R. (2000): A
comparative analysis of current credit risk models. {\em Journal
of Banking and Finance} 24, 59-117.

\bibitem{dorfleitner} Dorfleitner G., Schneider P. and Veza T. (2007): Flexing the Default Barrier. Working paper. Available at {\tt ssrn.com}.


\bibitem{Giesecke2002} Giesecke K. (2002): Correlated default with
incomplete information. {\em Cornell University working paper}, to
appear in {\em J. of Banking and Finance}.

\bibitem{Giesecke2004} Giesecke K. (2004): Credit Modelling and Valuation. An Introduction.
To appear in {\em Credit Risk: Models and Management, Vol. 2, Riskbooks, London.}.

\bibitem{hsu} Hsu, J. C., Sa´a-Requejo, J., and Santa-Clara, P. (2002). Bond pricing with default risk. Working paper, Anderson School, UCLA.

\bibitem{Huge} Huge B. and Lando D. (1999): Swap Pricing with Two-Sided Default Risk in a Rating-Based Model, {\em European Finance Review} 3, 239-268.

\bibitem{hui} Hui, C. H., Lo, C. F., and Tsang, S. W. (2003). Pricing corporate bonds with dynamic default barriers. {\em Journal of Risk}, 5(3).


\bibitem{Leung} Leung S. Y. and Kwok Y. K. (2005): Credit Default Swap Valuation with Counterparty Risk. {\em The Kyoto Economic Review} 74 (1), 25-45.

\bibitem{lipton} Lipton A. and Sepp A. (2009): Credit value adjustment for credit default swaps via the structural default model. {\em Journal of Credit Risk} Vol. 5, N. 2, 123-146.

\bibitem{LLH} Lo C. F., Lee H. C. and Hui C. H. (2003):
A Simple Approach for Pricing Barrier Options with
Time-dependent Parameters. {\em Quantitative Finance} 3.

\bibitem{merton74} Merton  R. (1974): On the pricing of corporate
debt: The risk structure of interest rates. {\em J. of Finance}
29, 449-470.


\bibitem{MP} Pignatelli M. (2004):
\emph{Private Communication,} Banca IMI.

\bibitem{Rapix} Rapisarda F. (2003):
Pricing Barriers on Underlyings with Time-dependent
Parameters. Working Paper.


\bibitem{Sorensen} Sorensen E. H. and Bollier T. F. (1994): Pricing Swap Default Risk, {\em Financial Analysts Journal} Vol. 50, N. 3., 23-33.

\bibitem{Walker} Walker M. (2005): Credit Default Swaps with Counterparty Risk: A Calibrated Markov Model. Working Paper.

\bibitem{Yi} Yi C. (2009): Dangerous Knowledge: Credit Value Adjustment with Credit Triggers. Bank of Montreal research paper.

\end{thebibliography}
\end{document}